\documentclass[12pt,preprint]{aastex}
\usepackage{color}
\usepackage{natbib}

\newcommand{\kvuma}{XTE J1118+480}

\begin{document}

\title{The Mass of the Black Hole in \kvuma}

\author{Juthika Khargharia}
\affil{SAS}
\affil{SAS Institute Inc., 820 SAS Campus Dr., Cary, NC, 27513}
\email{juthikak@gmail.com}

\author{Cynthia S. Froning\altaffilmark{1}}
\affil{Astrophysical and Planetary Sciences}
\affil{University of Colorado, 391, UCB \\ Boulder, CO 80309}
\email{cynthia.froning@colorado.edu}

\author{Edward L. Robinson}
\affil{Department of Astronomy}
\affil{University of Texas at Austin, Austin, TX 78712}
\email{elr@astro.as.utexas.edu}

\and

\author{Dawn M. Gelino}
\affil{NASA Exoplanet Science Institute}
\affil{Caltech, 770 South Wilson Avenue, Pasadena , CA 91125}
\email{dawn@ipac.caltech.edu}

\altaffiltext{1}{Center for Astrophysics and Space Astronomy, University of Colorado, 593, UCB, Boulder, CO 80309-0593}

\begin{abstract}

We present contemporaneous, broadband, near-infrared spectroscopy (0.9 -- 2.45 $\mu$m) and H-band photometry of the black hole X-ray binary, \kvuma. We determined the fractional dilution of the NIR ellipsoidal light curves of the donor star from other emission sources in the system by comparing the absorption features in the spectrum  with field stars of known spectral type.  We constrained the donor star spectral type to K7 V -- M1 V and determined that the donor star contributed 54$\pm$27\% of the H-band flux at the epoch of our observations. This result underscores the conclusion that the donor star cannot be assumed to be the only NIR emission source in quiescent X-ray binaries. The H-band light curve shows a double-humped asymmetric modulation with extra flux at orbital phase 0.75. The light curve was fit with a donor star model light curve, taking into account a constant second flux component based on the dilution analysis. We also fit models that included emission from the donor star, a constant component from the accretion disk, and a phase-variable component from the bright spot where the mass accretion stream impacts the disk. These simple models with reasonable estimates for the component physical parameters can fully account for the observed light curve, including the extra emission at phase 0.75. From our fits, we constrained the binary inclination to $68^\circ \leq i \leq 79^\circ$. This leads to a black hole mass of $6.9 M_{\sun} \leq M_{BH} \leq 8.2 M_{\sun}$.  Long-term variations in the NIR light curve shape in \kvuma\ are similar to those seen in other X-ray binaries and demonstrate the presence of continued activity and variability in these systems even when in full quiescence.

\end{abstract}


\section{Introduction}

\kvuma\ belongs to the class of transient low mass X-ray binaries (LMXBs) in which a late-type donor star transfers mass to its compact companion (either a black hole or a neutron star) through an accretion disk. It was discovered by the All-Sky Monitor (ASM) aboard the Rossi X-ray Timing Explorer (RXTE) satellite when it went into outburst in 2000 \citep{remillard2000}. Because of its location along a sightline of low interstellar absorption (E[B--V] = 0.013), this system has been extensively studied both in outburst and in quiescence at multiple wavelengths \citep[e.g.,][]{hynes2000,mcclintock2001,fender2001,chaty2003}. Observations obtained during outburst and after the system had returned to quiescence established several dynamical properties of the binary, including its orbital period ($P_{orb}=0.17$ days), donor star radial velocity semi-amplitude (K$_{2}$ = 709 km s$^{-1}$), and a mass function of the compact accretor of $f(M)=6.27\pm.04$$M_{\odot}$  \citep{mcclintock2001b, wagner2001,gonzalez2008}. The mass function is the minimum mass of the compact object, establishing that \kvuma\ contains a black hole. 

Accurate compact object masses are required to test models of formation and evolution of black holes and neutron stars \citep{brown1998, fryer2001,nelemans2001, belczynski2012}.  A black hole is characterized by its mass and spin, and a reliable estimate of the latter is intricately dependent on the former \citep[e.g.,][]{steiner2009}. These two parameters are critical in understanding space-time behavior near a black hole \citep{mcclintock2011}. X-ray binaries with black hole accretors have been used to test accretion disk and jet models \citep{yuan2005,maitra2009}, to test models of natal kicks and spin-orbit misalignments between the black hole and the accretion disk \citep{fragos2009}, and to predict observational tests of braneworld  gravity models \citep{johannsen2009}.  Observations of black holes in X-ray binaries are also important for understanding supermassive black holes in AGNs. Studies have revealed a ``Fundamental Plane'' of black hole accretion in which X-ray luminosity, radio luminosity, and black mass are related, both for Galactic black holes and their supermassive counterparts, implying that physical processes in AGNs can be illuminated by observations of similar processes in X-ray binaries by an appropriate scaling of the black hole mass \citep{merloni2003,falcke2004}. 

Compact object mass determinations require values for the binary orbital period, the radial velocity semi-amplitude of the donor star, the mass ratio, and the inclination. The latter is usually determined by modeling the ellipsoidal modulation of the light curve caused by the Roche lobe-filling donor star. The light curve observations are often performed in the near-infrared (NIR), where the late-type donor star dominates.  Many investigators have assumed a negligible contribution from non-stellar sources (e.g., the accretion disk and/or a jet outflow) at NIR wavelengths. However this assumption has been questioned based on fits to the NIR spectra with donor star templates and on variability in the NIR light curves during quiescence \citep{froning2007,reynolds2008,cantrell2008, cantrell2010}. Unaccounted-for extra flux will lower the derived binary inclination from its true value and consequently overestimate the compact object mass, so accurate determinations of the relative contributions of the donor star and other emitters in the NIR are necessary to obtain correct mass values. In a recent study of the effects of systematic errors on the derived masses of black holes in X-ray transients, \citet{kreidberg2012} showed that the assumption of zero non-stellar light biases derived black hole masses to higher than their true values, often by significant amounts. They also cautioned that the non-stellar component often varies over time and with orbital phase, requiring careful modeling of the observed light curve and contemporaneous acquisition of light curve and spectroscopic data to obtain unbiased compact object masses. 

\citet{gelino2006} obtained the most comprehensive light curve data set for \kvuma, covering the B, V,  R,  J, H and K wavebands.  They simultaneously modeled all the light curves to derive an inclination of $68^{+2.8}_{-2.0}$ degrees and a black hole mass of $M_{BH} = 8.53\pm0.6$~M$_{\sun}$. However, they assumed the non-stellar contribution in the NIR was negligible, $<$8\% dilution of the donor star light. Their inclination value was lower than the values found by other investigators (ranging from $71^{\circ}$ to $82^{\circ}$) using a variety of analysis methods, which may be due to Gelino et al.'s assumption of negligible non-stellar flux in the NIR \citep{wagner2001,zurita2002,mikolajewska2005,khruzina2005}. Changes in the shape of the NIR light curve over short time periods ---  the J-band light curve taken by \citet{mikolajewska2005} shows uneven peaks, unlike the symmetric ones seen by \citet{gelino2006} five months earlier --- suggests that variable, non-stellar emission is in fact present in \kvuma\ in the NIR.  
 
Motivated by the goal of determining an accurate black hole mass in \kvuma, we obtained contemporaneous light curve and spectral data of this system at NIR wavelengths to establish the non-stellar dilution from spectroscopy at the time of acquisition of light curve so that true binary inclination of the system can be obtained. In the following sections, we estimate the veiling caused by the non-stellar components at NIR wavelengths by examining the broadband spectral energy distribution and by measuring the equivalent widths of absorption lines in the photosphere of the donor star compared to field stars of known spectral type. We account for the extra NIR light when fitting ellipsoidal models to the observed light curve data in order to obtain a robust value of binary inclination and consequently determine the black hole mass. 

\section{Observations}\label{section:observations}

We obtained contemporaneous spectroscopic and photometric observations of \kvuma\ using the the Gemini Near-Infrared Spectrograph (GNIRS) at Gemini-North and the Near-Infrared Camera \& Fabry-Perot Spectrometer (NICFPS) on the 3.5-m telescope at Apache Point Observatory (APO). Photometry and spectroscopy were initially scheduled to be acquired on 2011 April 2 and 3 at both locations. We observed successfully at both sites on April 2. Additional light curve data were obtained on April 3 but due to bad weather at Mauna Kea, we did not complete the spectral observations until April 12. Table~\ref{table:xteobservations} lists the observations, exposure times, and  orbital phase coverage for the program.

\subsection{Spectroscopy}\label{section:observations:spectroscopy}

\begin{figure}
\figurenum{1}
\epsscale{1.0}
\plotone{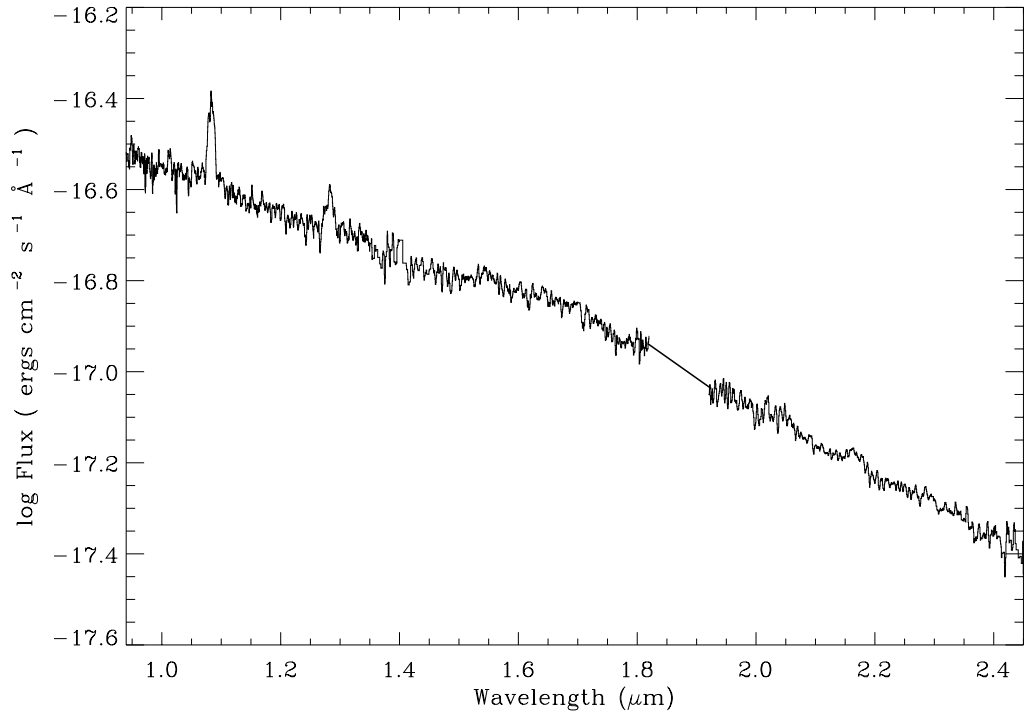}
\caption{The time-averaged spectrum of \kvuma\ obtained after correcting for atmospheric absorption and shifting the individual exposures to the rest frame of the donor star. The dereddened spectrum is shown. the SpeX short cross-dispersed observing mode does not cover the wavelength region from 1.86 -- 1.93~$\mu$m, which is indicated by a straight line in the figure.}
\label{fig:xtespectrum}
\end{figure}

We observed \kvuma\ using GNIRS at Gemini-North \citep{elias2006}.  GNIRS was configured in the cross-dispersed mode with the 31.7 line mm$^{-1}$ grating and the 0.30$\arcsec$ wide slit, yielding R $\sim$1700 while effectively covering the near-infrared range from 0.9 -- 2.5 $\mu$m. We obtained a total on-source observing time of 4.2 hrs. Data were taken in ABBA pairs at two positions along the 7.0$\arcsec$-long slit with a 310 sec exposure time at each position. The slit was oriented along the mean parallactic angle throughout the observations. An A0 V type telluric star was observed hourly with the same configuration. From two nights of data we were able to cover all binary phases at least once, except phases $\phi =$ 0.10 -- 0.35 and $\phi =$ 0.72 -- 0.76, which were covered twice. The raw data were reduced using the GNIRS tools within the Gemini-IRAF package, version v1.11. The raw images were first treated for fixed pattern noise using the {\tt cleanir}\footnote{http://staff.gemini.edu/~astephens/niri/cleanir/} tool. The other data reduction steps consisted of flat-fielding, sky subtraction and wavelength calibration. Owing to the faintness of the target, we could not confidently extract the spectra from the individual sky-subtracted exposures. Instead, we combined every three exposures (adjacent in phase and equivalent to 5\% of one full orbit) to perform reliable spectral extraction. The spectra extracted in this manner were minimally affected by orbital smearing effects when compared with the instrumental resolution of 177 km~s$^{-1}$.  Flux calibration and telluric correction were performed with the {\tt xtellcor} package within Spextool (version 3.4) \citep{cushing2004, vacca2003}. We shifted each of the extracted spectra to the rest frame of the donor star using the orbital ephemeris of \citet{calvelo2009}, median-combined each order separately, and merged all the orders to generate the time-averaged spectrum of \kvuma, which is shown in  Figure~\ref{fig:xtespectrum}. The spectrum has been boxcar-smoothed by 2 pixels, corresponding to one resolution element. The spectrum in  Figure~\ref{fig:xtespectrum} was corrected for interstellar absorption using A$_{V}$=0.066 \citep{gelino2006}. We did not attempt to quantify the flux calibration accuracy since the spectral analysis of \kvuma\ does not require absolute flux values. However, we find that the mean H-band flux of the spectrum agrees with the photometry within $\sim$ 4\%. 

\subsection{Photometry}\label{section:observations:photometry}

We obtained H-band light curve data with NICFPS at APO \citep{hearty2005}. We used individual exposure times of 20 seconds for the target. A standard star for flux calibration was observed every 1.5 hrs. We had excellent seeing conditions ($\sim0.6\arcsec$) on the night of April 2, but on April 3 the seeing conditions went from moderate ($\sim0.8\arcsec$) to poor ($>1.0\arcsec$) after the first two hours of observing and the data were discarded. The binary orbital phases between $\phi$ = 0.1--0.2 and $\phi$  = 0.33--0.5 were not covered at all, but all other phases were covered more than once. To account for the dark current in the NICFPS exposures, dark frames matching  the exposure times of the target and standard star were obtained at the beginning of each observing run and subtracted from the target /standard star exposures. Finally, the images were flat-fielded with sky-flats. We obtained five point dither images of the field, each offset by 20$\arcsec$, and constructed a sky-flat by median-combining these exposures. To obtain sufficient signal to noise in our reduced data images, we median combined every three exposures (equivalent to $\sim$ 60 seconds) to construct final reduced images. Aperture photometry was performed on the combined images using the IRAF {\tt phot} package. For flux calibration, we used the stars in the field of \kvuma\ selected from the Two Micron Sky Survey (2MASS) catalogue as well as the ARNICA (Arcetri NICMOS3 camera) near-infrared standard star AS-11 \citep{hunt1998}. 

\section{Analysis}

\subsection{The Donor Star Spectral Type and Fractional Contribution to the NIR Spectrum}

\begin{figure}
\figurenum{2}
\epsscale{1.0}
\plotone{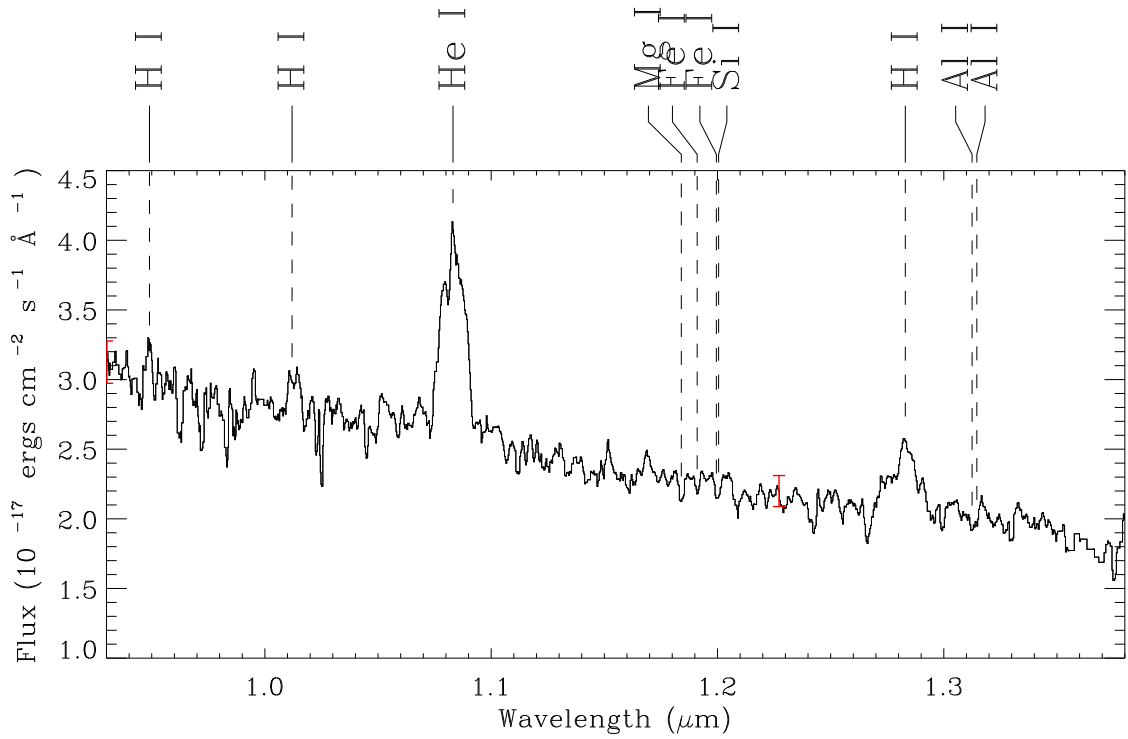}
\caption{The J-band spectrum of \kvuma\ showing emission features from the accretion disk and absorption features from the donor star. The typical error bar is shown in red.}
\label{fig:xtejband}
\end{figure}

\begin{figure}
\figurenum{3}
\epsscale{1.0}
\plotone{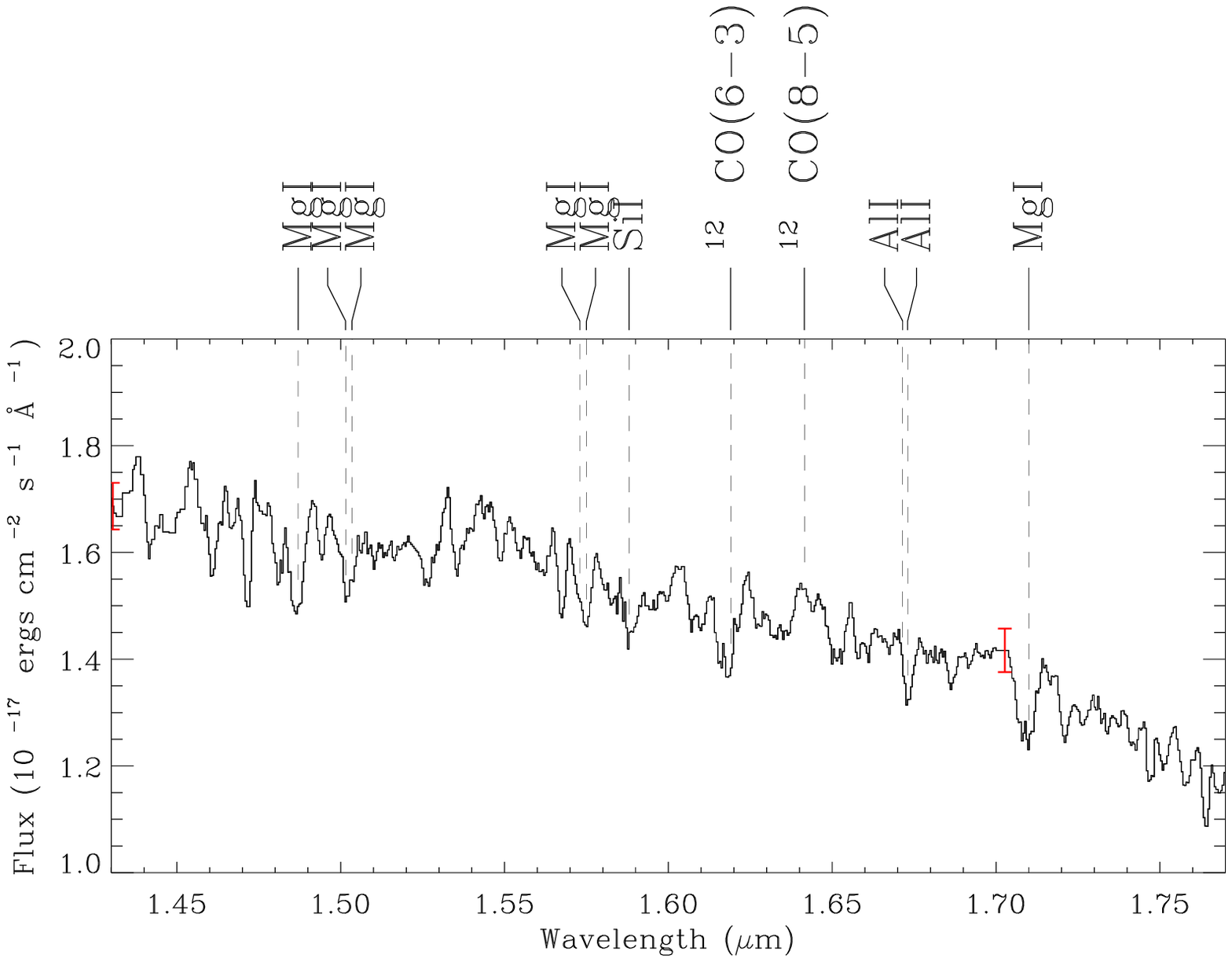}
\label{fig:xtehband}
\caption{The H-band spectrum of \kvuma.}
\end{figure}

\begin{figure}
\figurenum{4}
\epsscale{1.0}
\plotone{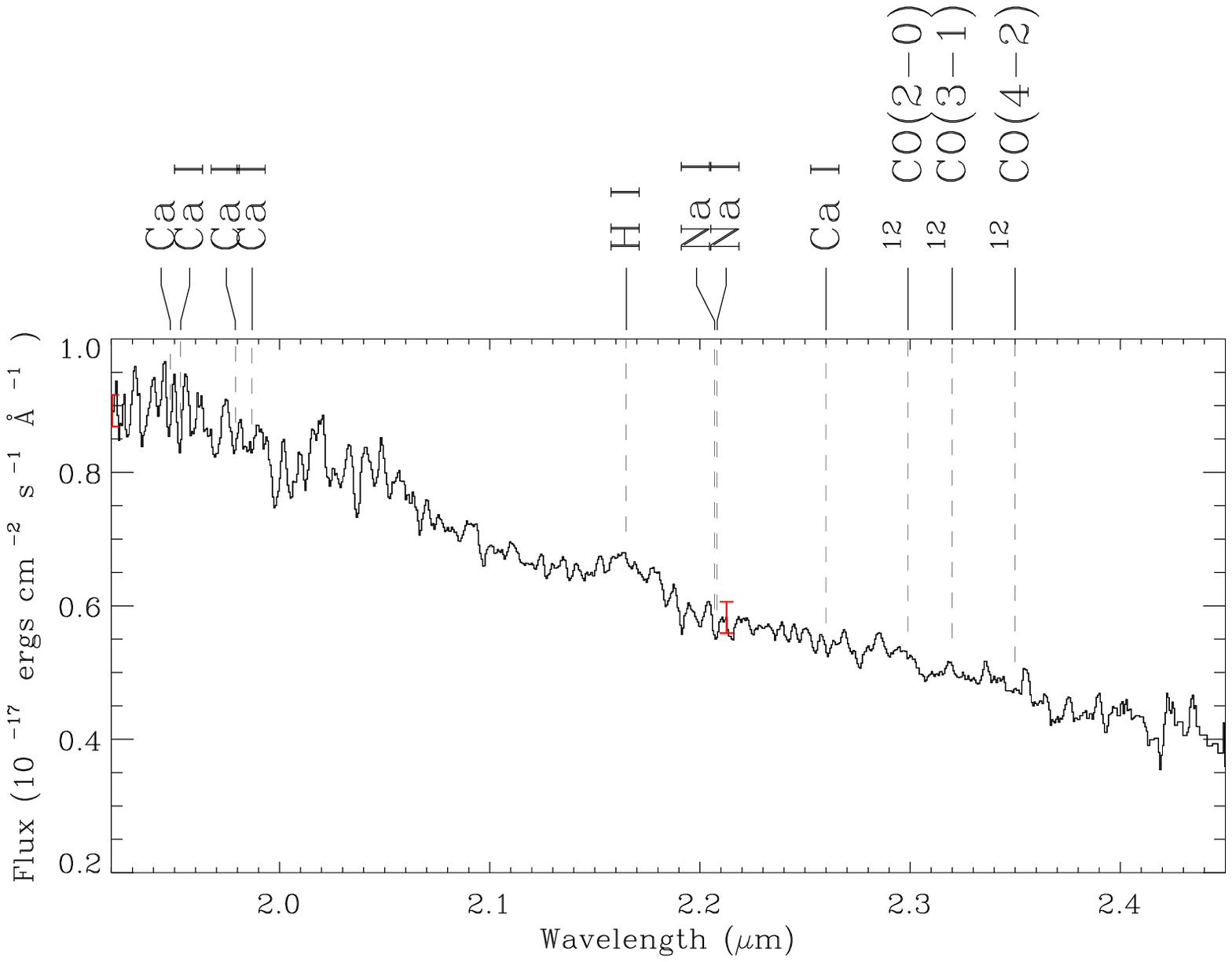}
\caption{The K-band spectrum of \kvuma.}
\label{fig:xtekband}
\end{figure}

The spectrum of \kvuma\ displays narrow absorption lines of neutral metals, including transitions of \ion{Al}{1}, \ion{Na}{1}, \ion{Mg}{1}, \ion{Fe}{1}, and \ion{Si}{1} that are believed to originate in the photosphere of the donor star. We also detect broad emission lines of \ion{H}{1} and \ion{He}{1} from the accretion disk. The J-, H- and K-band spectra are shown in Figures~\ref{fig:xtejband}, \ref{fig:xtehband} and \ref{fig:xtekband}. The error bars shown were calculated for each resolution element (2 pixels) by fitting straight lines through several continuum-dominated regions and obtaining the rms scatter about the fit. The continuum-dominated regions were selected after an inspection of similar regions in K5 V -- M1 V template stars. We obtained signal-to-noise (S/N) estimates of $\sim18$ in the H-band and $\sim15$ in the J- and K-bands, respectively. 

In the J-band we detect a few absorption features in the 1.175 -- 1.320 $\mu$m range, mainly blends of \ion{Mg}{1}, \ion{Si}{1}, \ion{Fe}{1}, and \ion{Al}{1}, and broad emission features of \ion{H}{1} and \ion{He}{1}. In the H-band, we observe narrow atomic features of \ion{Mg}{1}, \ion{Al}{1}, and \ion{Si}{1}. In the K-band, the most prominent features used for spectral classification are the $^{12}$CO bands, \ion{Na}{1}, and \ion{Ca}{1}, with \ion{Na}{1}. The position of these lines are marked in Figure~\ref{fig:xtekband}. There is also a weak emission feature of \ion{H}{1} at 2.16~$\mu$m. Most of the absorption features are detected at low confidence or undetected in K.  We do not detect any CO features in the K-band. Enhanced \ion{N}{5} and depleted \ion{C}{4} and \ion{O}{5} emission lines have been seen in the UV spectrum of \kvuma, suggesting that the accreting material has been CNO processed \citep{haswell2002}. Therefore, our non-detection of the CO-bands in the K-band is unsurprising. In the H-band, we apparently detect the $^{12}$CO (6,3) feature at 1.619 $\mu$m, but given the relative oscillator strengths of the K- and H-band lines (the second overtone, $\Delta\nu = 3$, bands of CO near the 1.1619 $\mu$m have oscillator strengths that are approximately 100 times smaller than the first overtone bands near 2.29 $\mu$m) and the presence of several other species contributing features in this wavelength region (including Ca, Fe, Ni, Si, and OH), we consider the apparent H-band CO feature spurious. 

\begin{figure}
\figurenum{5}
\epsscale{1.0}
\plotone{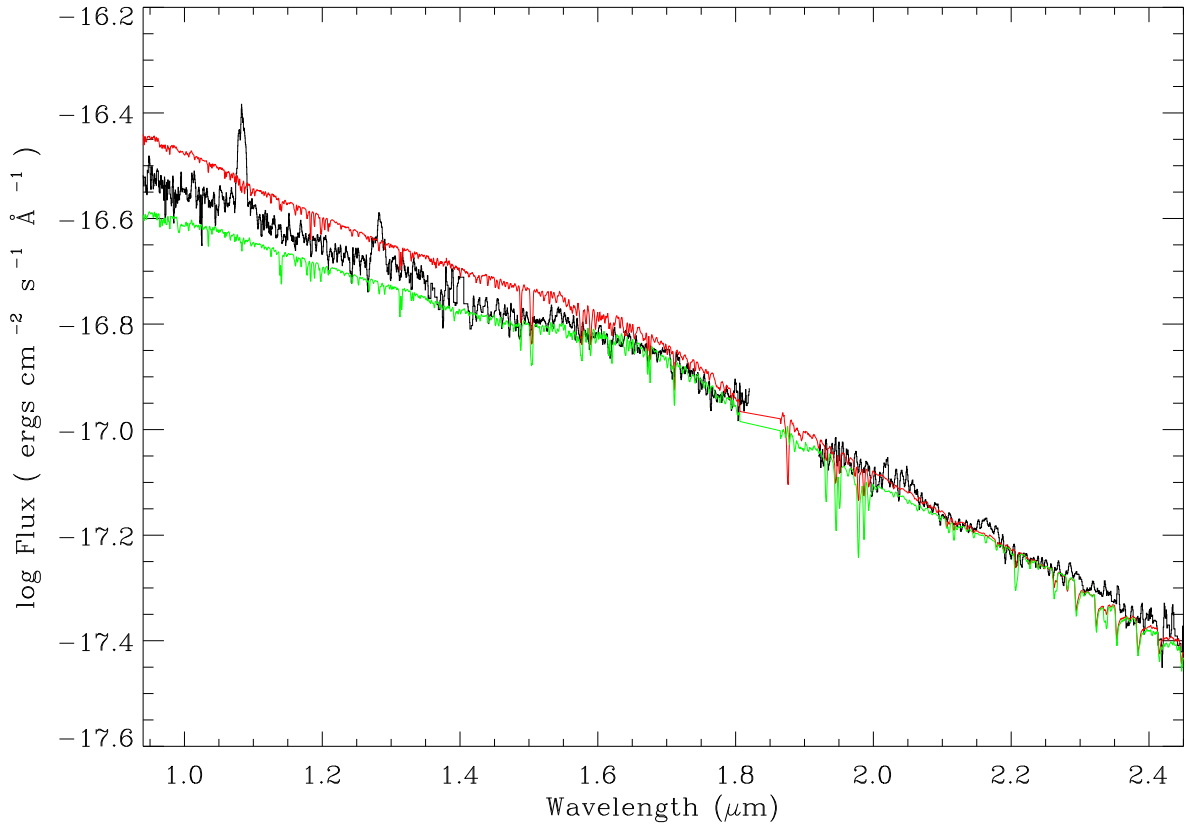}
\caption{The time-averaged spectrum of \kvuma\ compared to the spectrum of a K5 V star (in red) and a M1 V star (in green). The K5 V and M1 V spectra have been scaled to match the \kvuma\ flux near 2.23 $\mu$m. }
\label{fig:xtecomparison}
\end{figure}

The spectral type of the donor star in \kvuma\ has been broadly classified as K5 V -- M1 V \citep{gonzalez2008, gelino2006, wagner2001, frontera2001, mcclintock2001}. To obtain a precise black hole mass, we want to constrain the spectral type of the donor star further and to estimate the relative contributions of NIR light from the donor star and other sources in the system. Towards that end, we examined the broadband spectral energy distribution (SED) of \kvuma\ compared to field stars. Figure~\ref{fig:xtecomparison} shows the spectrum of \kvuma\ compared to the spectrum of a K5 V star (in red) and an M1 V star (in green). The broad H-band bump (between 1.5--1.7 $\mu$m) that is seen in the spectra of K-stars increases in amplitude with decreasing effective temperature and is attributed to the H$^{-}$ opacity minimum at 1.6~$\mu$m \citep{rayner2009}. In \kvuma, the size of this feature is consistent with a donor star spectral type later than K5 V. 

Earlier studies assumed that the donor star is the sole source of NIR emission in \kvuma. If we follow that assumption and scale the field star spectra to match the observed flux in K (at 2.23~$\mu$m), we find that the template field star spectrum unphysically exceeds the observed spectrum at shorter wavelengths ($>$7\% in both J and H) for K7 V or earlier spectral types. Thus, some dilution of the donor star flux is required if the spectral type is K7 V or earlier. Based solely on the broad spectral shape, an MI V or later donor spectral type could be responsible for most or all of the NIR flux, at least for wavelengths $>1.5 \mu$m.

To constrain the donor spectral type further and estimate the fraction of NIR flux originating from the donor star (donor fraction, $f$) in \kvuma, we followed the procedure outlined in \citet{khargharia2010} and \citet{froning2007}. Specifically, we fit normalized template spectra of known spectral type to the normalized spectrum of \kvuma. Dilution of the X-ray binary spectrum by non-stellar emission sources will have the effect of decreasing the equivalent widths of the lines in the observed spectrum compared to those in the templates. By scaling the templates to fit the observed line strengths, we can determine the donor star fractional contribution to the NIR spectrum of \kvuma. 

To normalize the spectra, we fit a spline function to the continuum, which was then divided out.  The continuum points were selected by eye. The normalized spectrum of the template star was scaled by a fraction $f$ that was varied from 0 to 1 in steps of 0.01. The normalized, scaled template was subtracted from the normalized spectrum of \kvuma\ and the residual was computed. The value of $f$ that minimized the deviation between the residual and the mean of the residual determined the best fit. For the template spectra, we used stars of spectral type K5 V (HD36003), K7 V (HD237903), M0/M0.5 V (HD209290) and M1 V (HD42581)  \citep[There is a disagreement in literature about the spectral type of HD209290 and hence we refer to it as M0/M0.5 V star][]{rayner2009, koen2010}. The template star spectra were obtained from the IRTF spectral library\footnote{\url{http://irtfweb.ifa.hawaii.edu/~spex/IRTF\_Spectral\_Library}}. 

The fit regions that were investigated in the J-, H- and K-bands for the spectral types between K5 V -- M1V are listed in Table~\ref{table:fieldstarfits2} along with the best fit donor fractions, $f$, and uncertainties on those values. In the J-band, our fits were restricted to the region between 1.20 -- 1.314 $\mu$m within which we detected features containing blends of \ion{Mg}{1}, \ion{Si}{1}, \ion{Fe}{1} and \ion{Al}{1}.  In the H-band, we fit features of \ion{Mg}{1}, \ion{Al}{1} and \ion{Si}{1} over the wavelength range 1.475 -- 1.73 $\mu$m. We only show fits in the K-band to the \ion{Na}{1} feature, although the detection of this line is marginal.

Since the S/N in our spectrum is low and the use of $\chi^2$ statistics for the spectral fitting are affected by systematic uncertainties \citep[see the discussion in][]{froning2007}, we have adopted a new technique to properly evaluate the noise associated with the spectrum of \kvuma\ and its contribution to the donor fractions. The procedure for this involves ``scrambling" the \kvuma\ spectrum and fitting the template star spectra to many such randomly ``scrambled" spectra in order to place robust error estimates on the best fit donor fractions. A detailed explanation can be found in the Appendix. The uncertainties calculated in this manner are shown in Table~\ref{table:fieldstarfits2}. 

We also examined if the calculated donor fractions were consistent with the overall shape of the broadband spectral energy distribution of \kvuma. To answer that question, we scaled the flux of each field star by the average donor fraction in the H-band  (near 1.6 $\mu$m) and computed the value by which the average flux of the scaled template spectrum drops below that of \kvuma\ in both the J- and K-bands. Less emphasis was given on obtaining a K-band match due to the fact that we only had the \ion{Na}{1} feature to compare with and this feature is at about the same level as the noise in the K-band. Depending on whether the corresponding drop in flux in the J-band was consistent with the average J-band donor fraction (within the bounds allowed by propagating the errors to calculate the average donor fraction) computed from Table~\ref{table:fieldstarfits2}, we could further constrain the donor spectral type. When a K5 V star was scaled to 43\% of the \kvuma\ flux at the center of the H-band, it resulted in an average flux drop of 41\% in the J-band (between 1.1 -- 1.3 $\mu$m) and 38\% in the K-band (near 2.20 $\mu$m), inconsistent with the average donor fractions expected from the fits to lines in those bands. A K7 V matches the drop in J-band flux but not in the K-band, while M0.5 V/M1 V matches both J and K-bands when their H-band fluxes were scaled to the corresponding average H-band donor fraction obtained from Table~\ref{table:fieldstarfits2}. This suggests that K7 -- M1 V is the most likely range of donor spectral types in \kvuma. This is also consistent with the shape of the broad H-band bump near 1.6~$\mu$m discussed above.

\begin{figure}
\figurenum{6}
\epsscale{1.0}
\plotone{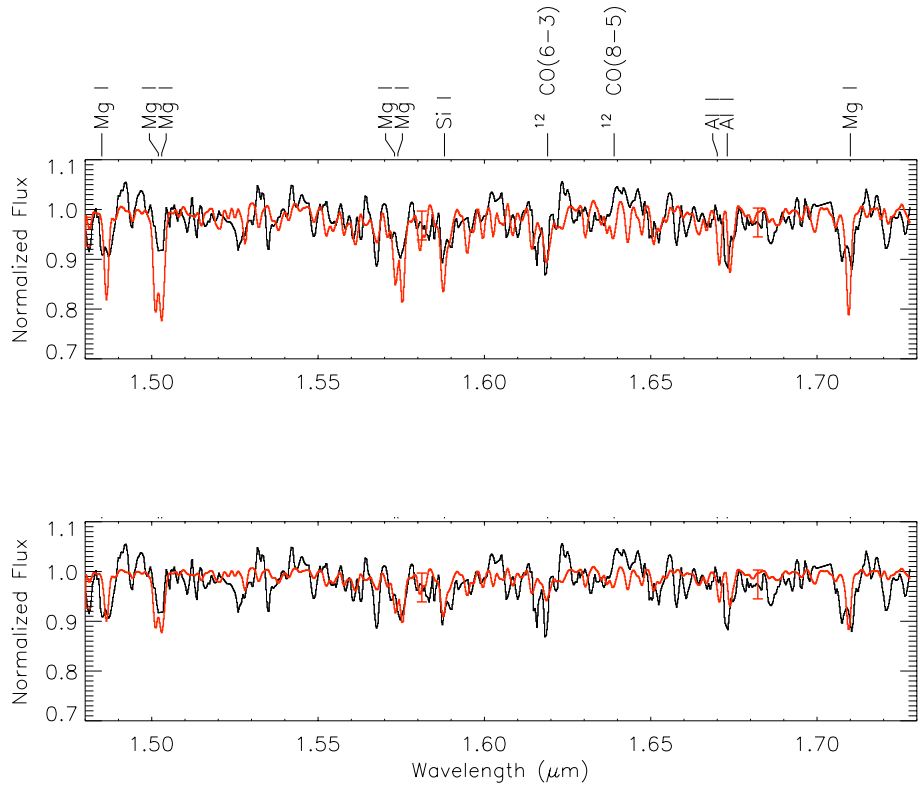}
\caption{The normalized H-band spectrum of \kvuma\ is shown in the upper panel with the normalized spectrum of a K7 V star overplotted (in red). The lower panel shows the same spectra after the K7 V star has been normalized to the best fit fractional contribution of 0.50.}
\label{fig:xtenormalk7}
\end{figure}

Thus, we conclude that the spectral type best describing the donor star in \kvuma\ lies between K7 V -- M1 V. We estimate the donor fraction by averaging the best fits to multiple lines using the K7 -- M1 V templates and propagating the corresponding uncertainties from Table~\ref{table:fieldstarfits2}. This leads to a H-band donor contribution of $f=0.50\pm0.32$ in the spectrum of \kvuma\ at the epoch of our observations. The upper panel of Figure~\ref{fig:xtenormalk7} shows the normalized spectrum of XTE J1118+480 plotted (in black) over the normalized spectrum of a K7 V star (in red) while the lower panel shows the same comparison but with the K7 V star scaled to match the best donor fraction of 50\%. 

\subsection{Modeling the Light Curve to Obtain the Inclination}

\begin{figure}
\figurenum{8}
\epsscale{1.0}
\plotone{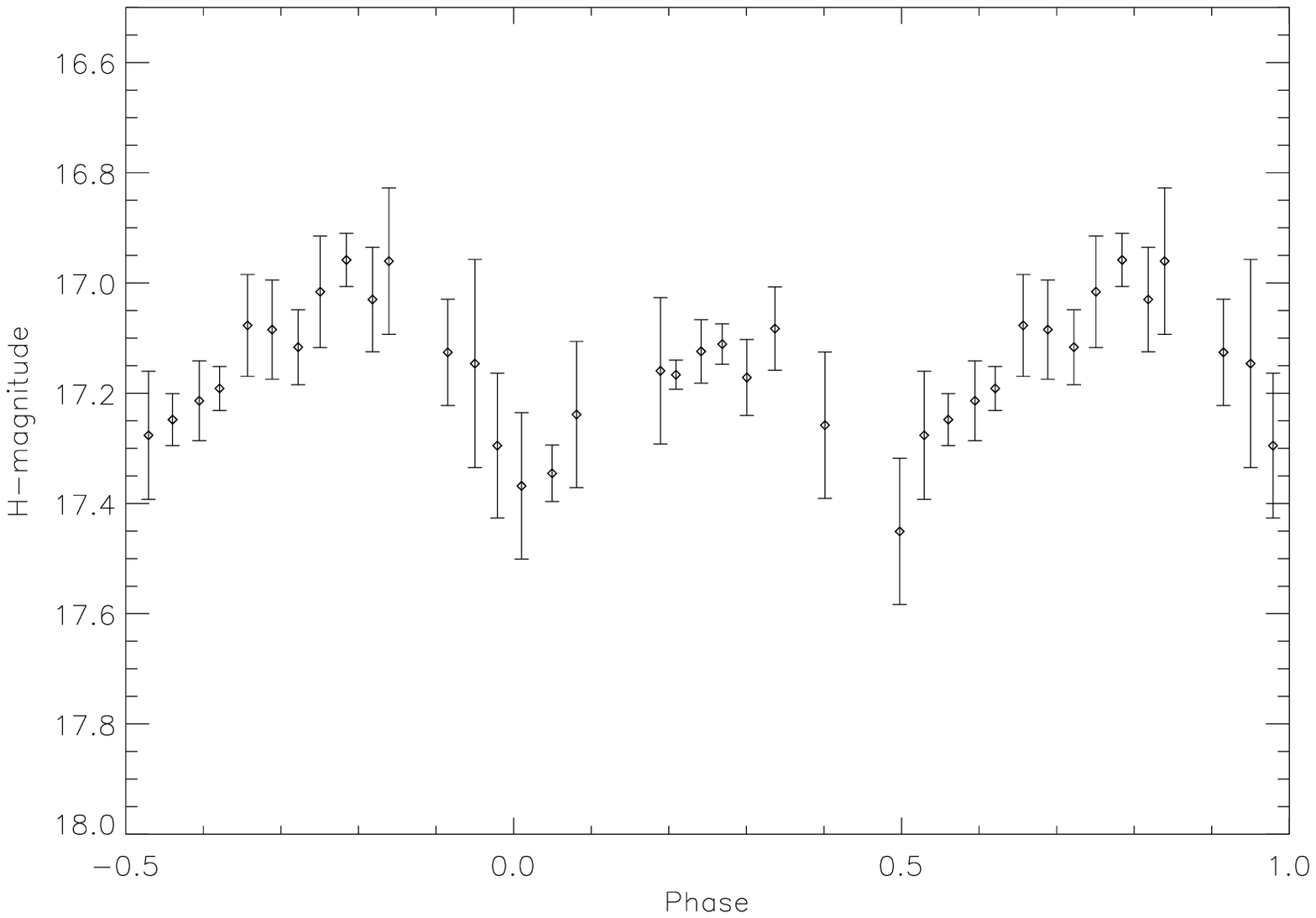}
\caption{The light curve of \kvuma\ after binning the observed data in phase bins of 0.03. The error bars are derived from scatter about the mean in each bin. For phase bins that contained fewer than 3 points, we used the largest error bar in the rest of the bins as the uncertainty.}
\label{fig:binnedlightcurve}
\end{figure}

Our spectroscopic observations of \kvuma\ were obtained in conjunction with contemporaneous light curve data. We phase-binned the H-band observations using the orbital ephemeris from \citet{gonzalez2008}. Figure~\ref{fig:binnedlightcurve} shows the H-band light curve after the data was combined into orbital phase bins of size $\Delta \phi=0.03$. Each datum in the figure represents the mean of the points in the specific phase bin. The error bars are the rms scatter of the points about the mean in each bin. In cases where we did not have enough data points ($<$3) to place a reliable error estimate on the magnitude, the error bar for that datum was changed to the value of the largest error bar for the other light curve points. The orbital phases represent the standard convention wherein phase 0.0 is inferior conjunction of the donor star. 

The H-band light curve in Figure~\ref{fig:binnedlightcurve} shows a departure from the conventional ellipsoidal modulation expected from a Roche-lobe filling donor star: the maximum at phase $\phi=0.75$ is higher than the maximum at $\phi=0.25$. Similar asymmetric modulations were detected in the quiescent J-band light curve of \kvuma\ obtained by \citet{mikolajewska2005}. However, \citet{gelino2006} obtained NIR light curves of \kvuma\ four months prior to \citet{mikolajewska2005} and found no asymmetries in their NIR light curves. Typically, this observed asymmetry in the light curve maxima is attributed to the emission from a bright spot in the accretion disk \citep{froning2001} or from a dark spot on the donor star \citep{gelino2001}. Evidence for the presence of a hot spot in \kvuma\ is ambiguous:  Doppler and modulation tomography of \kvuma\ taken during quiescence by \citet{calvelo2009} revealed a well-defined hotspot, but \citet{torres2004} found no evidence of a hotspot in their Doppler tomograms. \citet{calvelo2009} speculated that variations in the mass transfer rate from the donor star during quiescence may cause the bright spot to be intermittent. 

In the following sections, we present two simple models for fitting the observed light curve of \kvuma: a) a model incorporating a donor star with constant extra flux from nonstellar sources (e.g., the accretion disk and/or a jet); and b) a model incorporating a donor star and an accretion disk with a bright spot. We modeled the H-band light curve of \kvuma\ using an updated version of the light curve synthesis code first presented in \cite{zhang1986} that has been used to model the light curves in several other LMXB's \citep{froning2001,khargharia2010}. The code accounts for the geometry of the binary system and then calculates the light curve by computing the temperature and intensity distribution across the surface of the Roche-lobe filling donor star and an accretion disk with a bright spot. It also takes into account gravity and limb darkening for each component of the system. The best fit light curve is obtained by minimizing the chi-squared value between the synthetic and the observed light curves. Below we discuss each model that was fit to the observed light curve of \kvuma. 

\subsubsection {Modeling the H-band light curve with a donor star and constant non-stellar flux}\label{section:modeldisk}

\begin{figure}
\figurenum{9}
\epsscale{1.0}
\plotone{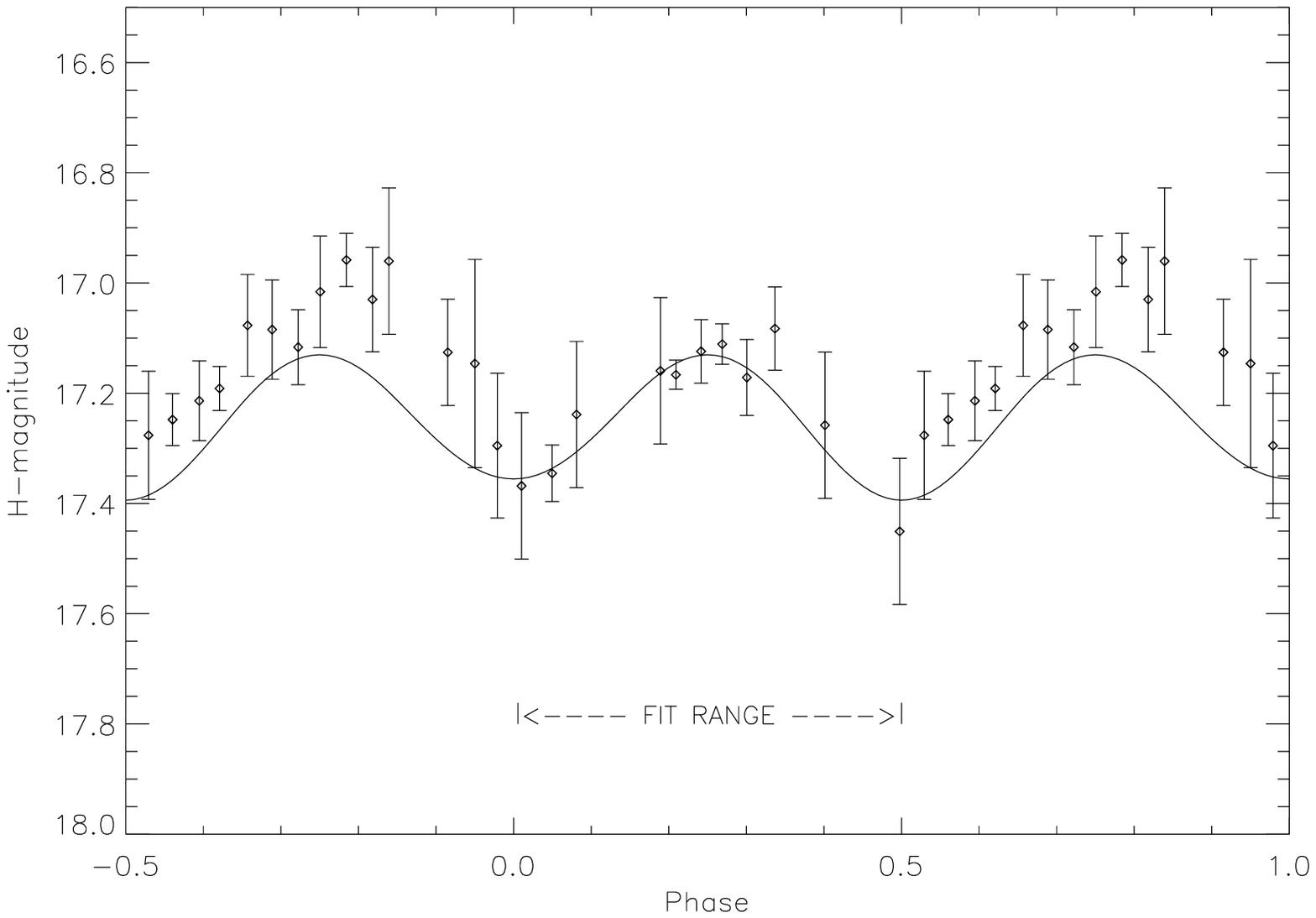}
\caption{The best fit light curve obtained by modeling a donor star (fit range: $\phi$= 0.0--0.5) with a constant extra flux component. The inclination is $68^\circ$.}
\label{fig:modeldisk}
\end{figure}

We ran models to evaluate the effect of adding constant (i.e., present over the full binary orbit) non-stellar flux on the inclination of the binary. For modeling purposes, the donor star parameters were set as follows: $T_{eff}=4000$~K, to reflect the average temperature from the spectral type range determined above; a gravity darkening coefficient of 0.08, assuming that the donor has a convective envelope \citep{lucy1967, sarna1989}; and limb-darkening coefficients from \citet{claret1995}. We adopted the mass ratio value obtained by \citet{calvelo2009}, which agrees with past estimates obtained by \citet{gonzalez2008}, \citet{torres2004} and \citet{zurita2002}. To avoid the extra flux at $\phi=0.75$, we modeled the light curve between phases $\phi=$0.0 -- 0.50 only. By allowing the constant non-stellar fraction ($1 - f$) to vary within the limits obtained from spectroscopy, we modeled the light curve data to derive the corresponding best fit binary inclinations. Using this method, we find the binary inclination to lie between 68$^\circ$ -- 89$^\circ$. Figure~\ref{fig:modeldisk} shows the best fit light curve ($i=68^\circ$) obtained by modeling the observed light curve with a donor star and for a disk light fraction of $f_{disk}=0.18$ ($\chi^2_{\nu}= 0.51$). We varied the mass ratio within the limits specified in \citet{calvelo2009} and the results were unchanged. From this, we conclude that accounting for the non-stellar flux within its uncertainty bounds obtained from spectroscopy places a lower limit on the binary inclination of $i \geq 68^\circ$. 

\subsubsection{ Modeling the H-band light curve with a donor star and an accretion disk with a bright spot}

The H-band light curve of XTE J1118+480 shows asymmetry in the maxima of the peaks at phases 0.25 and 0.75, which is often attributed to emission from a bright spot on the accretion disk. Accordingly, we also fit the light curve with a model that includes a donor star and a cool opaque accretion disk with a bright spot on its rim. By this method, we do not attempt to constrain the physical properties associated with either the disk or the bright spot. (Indeed, there is no a priori reason to believe that the non-variable emission comes from the accretion disk rather than partially or wholly from a persistent jet.) However, we can derive a range of possible inclinations by modeling this system with reasonable estimates for the parameters associated with the accretion disk and bright spot. 

The donor star parameters were set at the same values used in the earlier models. Varying the donor star parameters had negligible effect ($\leq 1^\circ$ change) on the derived inclination of the system and therefore we keep them fixed for these set of models and only vary the parameters associated with the accretion disk and the bright spot. For the accretion disk, we adopted an inner radius of $0.001R_{L_{1}}$ (where $R_{L_{1}}$ represents the distance to the inner Lagrangian point) and an the outer disk radius of $0.75R_{L_{1}}$; these numbers are taken from the work done by \citet{wren2001} and \citet{calvelo2009}. We set the flare half-angle of the disk to a small value of $1^{\circ}$. The bright spot was added to the rim of the disk and spans an angle in the azimuthal direction whose position $\phi_{spot}$ and width $\Delta \phi_{spot}$ are variable parameters in the light curve fitting model. The other input parameters are  the temperature of the disk ($T_{disk}$) and the bright spot ($T_{spot}$). Both the accretion disk and the bright spot were assumed to emit as black bodies with single temperatures and linear limb-darkening coefficients.  
We varied $T_{disk}$ between $2500 - 4500$ $K$, $T_{spot}$ from $5000 - 20000$ $K$, $\phi_{spot}$ from $40^{\circ}$ -- 120$^{\circ}$, and $\Delta\phi_{spot}$ from $5^{\circ}$ -- 15$^{\circ}$. The choice of $T_{disk}$ was motivated by the work done in \citet{reynolds2008}, who found that thermal emission from a cool outer accretion disk ($T_{disk}\sim$ 2000 -- 4000~K) could be used to model the excess NIR emission in the multi-wavelength SED of several XRBs (including \kvuma). 

\begin{figure}
\figurenum{10}
\epsscale{1.0}
\plotone{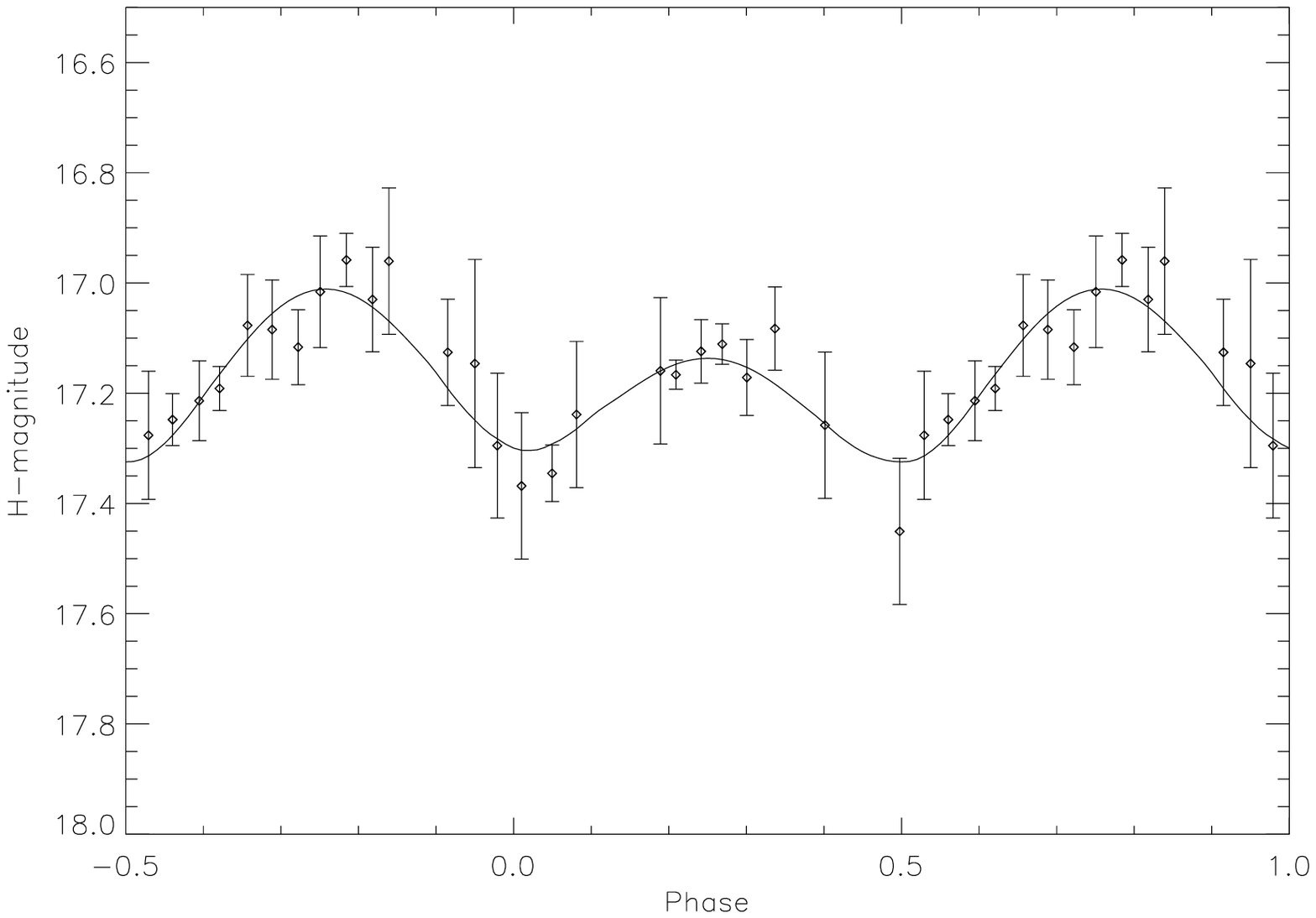}
\caption{Light curve models consisting of a donor star and an accretion disk along with a bright spot are fit to the full light curve of \kvuma. The best fit parameters for this model are $i=78^\circ$, $T_{disk}=3000K$, $T_{spot}=12000K$, $\phi_{spot}=85^\circ$, $\Delta\phi_{spot}=5^\circ$}
\label{fig:modelbrightspot}
\end{figure}

Using these parameters, we fit models to the full H-band light curve. In Figure~\ref{fig:modelbrightspot}, we show the best fit light curve ($\chi^2_{\nu}$= 0.57), with an inclination of $78^\circ$, $T_{disk}=3000$ $K$, $T_{spot}=12000$ $K$, $\phi_{spot}=85^\circ$ and $\Delta\phi_{spot}=5^\circ$. From an investigation of our fits over the entire range of input parameters, we find that all fits have $\chi^2_{\nu}$ values between 0.60 -- 0.80 and produced lower/higher $T_{disk}$ values for correspondingly lower/higher $T_{spot}$ values; e.g., a model with $T_{disk}=2500$ $K$ and $T_{spot}=10000$ $K$ for an inclination of $i=74^\circ$ gave a comparable fit ($\chi^2_{\nu}$ =0.60) to a model with $T_{disk}=4000$ $K$ and $T_{spot}=13000$ $K$ for an inclination for $i=81^\circ$ ($\chi^2_{\nu}$ = 0.63). 

We do not attempt to set constraints on either the disk or the bright spot parameters based on the light curve models with the lowest $\chi^2_{\nu}$ value, except to note that a simple model incorporating a donor star and an accretion disk with a bright spot on the rim can account for the extra flux at $\phi=0.75$. More important, we can set an upper limit on the binary inclination through the absence of eclipse features in the light curve. Extensive multiwavelength observations of \kvuma\ in outburst found no evidence of eclipses in this system \citep{uemura2000,wood2000}. From our modeling, we find that eclipse features start to emerge in the model light curves at $i \geq 80^\circ$ for an accretion disk of outer radius $0.75R_{L}$. Hence, to maintain consistency with the non-detection of eclipses in \kvuma, we find an upper limit to the binary inclination of $\leq79^\circ$.

In modeling asymmetric light curve peaks in X-ray binaries, some authors have attributed the asymmetry to a dark spot on the donor star rather than a bright spot on the accretion disk \citep{gelino2001}. In that case, our assumption in the previous section that modeling orbital phases  $\phi=$0.0 -- 0.50 is preferable to modeling the full light curve is exactly backwards and we would have underestimated the true amplitude of the ellipsoidal modulation (as we would then be modeling the hump artificially depressed by the stellar spot).  If so, our lower limit to the inclination of $\geq 68^{\circ}$ is conservative but still encompasses the higher inclination that would result from fitting the larger hump in the light curve.

In summary, by modeling the observed light curve with a constant non-stellar fraction obtained from spectroscopy, we found a lower limit to the binary inclination in \kvuma\ as $\geq68^\circ$. Additionally, an upper limit on the inclination is dictated by the absence of eclipses, as $<80^\circ$. Hence, we have constrained the binary inclination to lie between $68^\circ \leq i \leq  79^\circ$.

\section{Discussion}

In Table~\ref{table:parameters}, we summarize the current best estimates for the physical parameters including the black hole mass for \kvuma. By performing spectroscopy of \kvuma, we find that the donor star contributes a fraction $f=0.50\pm0.32$ of the H-band flux. By varying the constant non-stellar fraction within the limits obtained from the spectroscopy while simultaneously imposing the absence of eclipse condition, we obtained a binary inclination of $68^\circ\leq i \leq 79^\circ$. Using this inclination range, combined with the orbital parameters from \citet{calvelo2009}, we find the mass of the black hole in \kvuma\ to be $6.9 \leq (M_{BH}/M_{\odot}) \leq 8.2$.  Previous analyses have found inclinations ranging from $i = 60^{\circ}$ to $i = 82^{\circ}$  \citep{frontera2001,zurita2002,mikolajewska2005,khruzina2005,gelino2006}. Unfortunately, because of the relatively low S/N of our spectrum, we were only able to constrain the donor star fraction broadly. Despite that, the binary inclination range we found was fairly narrow ($12^\circ$) and we were able to provide a new constraint on the mass of the black hole in \kvuma.  Interestingly, the binary inclination of $68^\circ\pm2^\circ$ determined by \citet{gelino2006} is at the low edge of our inclination range, suggesting that they may have acquired their light curve data when the NIR non-stellar contribution was small.  

In an extensive study of the X-ray binary A0620-00, \citet{cantrell2008, cantrell2010} showed that the system existed in three distinct optical states even in quiescence. The authors found that a correct determination of the inclination relies on identifying the state of the system. This was found to be more important than the particular waveband where the measurements are made since non-stellar sources were present at both optical and NIR wavelengths. \kvuma\ also exhibits changes in the shape of its quiescent ellipsoidal light curve (e.g., comparing the data of \citealt{gelino2006} and \citealt{mikolajewska2005}). The asymmetry we observed in the H-band light curve of \kvuma\ is similar to that seen by \citet{mikolajewska2005} in their J-band data. Significant changes in the quiescent infrared light curves of the black hole X-ray binary GRO J0422+32 were also seen in two independent observations by \citet{reynolds2007} and \citet{gelino2003} even though the mean K-band magnitudes did not vary significantly.  From these studies, is clear that X-ray binary systems continue to harbor extensive activity and multiple accretion and/or outflow states even when fully in quiescence.

\section{Conclusions} 

We have obtained broadband near-infrared spectroscopy of \kvuma\ and contemporaneous light curve data to accurately account for the veiling that affects determinations of the binary inclination and compact object mass. By comparing the shape of the spectral energy distribution of the combined NIR spectrum as well as individual absorption lines in our spectrum with those of field stars of known spectral type, we were able to broadly account for the fraction of NIR light contributed by the donor star as $f=0.50\pm0.32$. We factored the non-stellar contribution into our H-band light curve fits and obtained a binary inclination of $68^\circ \leq i \leq79^\circ$. From these we obtained a robust determination of the black hole mass in \kvuma\ of $6.9 < (M_{BH}/M_{\odot}) \leq 8.2$. Our results are consistent with the picture of continued activity in X-ray binary systems even in quiescence and underscore the importance of accounting for all emission sources in the binary system when modeling spectra and light curves, even at NIR wavelengths. 
 
\acknowledgments{We would like to thank Emma Hogan for her assistance with Gemini-IRAF and Bernadette Rogers for letting us use her code to make GNIRS data compatible for use with Spextool}.

Facilities: \facility{Gemini-N (GNIRS), APO (NICFPS)}

\appendix 

\section{Determination of error in the donor fraction calculation}

Due to the low SNR of our spectrum, the donor fraction obtained from the spectrum of \kvuma\  is only meaningful if accompanied by reliable uncertainty estimates. In order to adequately account for the noise in the spectrum, we created ``scrambled'' versions of the observed spectrum and used fits to the noise spectra to determine the uncertainties on the donor star fraction. For each of the J-, H-, and K-band spectra, we obtained the Fourier transform of the normalized spectrum. The Fourier-transformed spectra give a complex valued array , $ X(k) = \sum\limits_{n=1}^{N} x(n) e^{{(-2\pi i/N)}(n-1) k}$, where N represents the total data points, $k$ represents the individual Fourier transformed frequency components, and x(n) represents the normalized flux values. The amplitude of the Fourier transform is given by $A(k) = \sqrt {Re [X(k)]^{2} + Im [X(k)]^{2}}$ and the phase is given by $\phi (k) = \tan^{-1}(-\frac{Im(X(k))}{Re(X(k))})$. 
 
\begin{figure}
\figurenum{11}
\epsscale{1.0}
\plotone{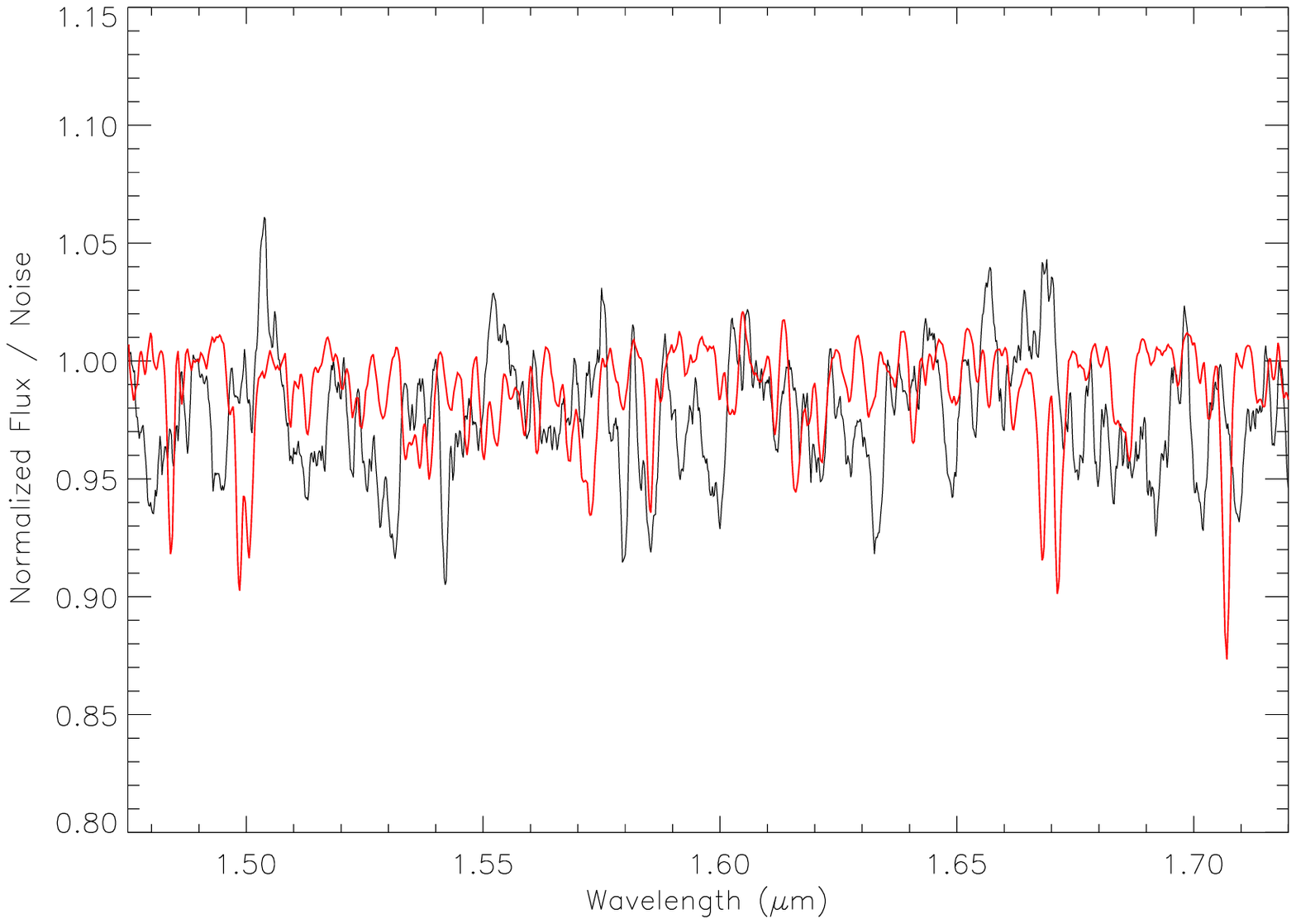}
\caption{An example of fitting an M0.5 V stellar spectrum to the inverse ``phase scrambled'' Fourier-transformed spectrum of \kvuma. The donor fraction variance obtained from fitting many such randomly phase-scrambled spectra were used to place uncertainties on the donor star fraction.}
\label{fig:fourier}
\end{figure}

To estimate the noise in our spectral data, we ``scrambled" the phase of the Fourier-transformed spectrum by adding a different random number between $0$ and $2\pi$ to the original phase at each $k$ while retaining the original power spectrum $|X_{k}|^2$. We used the {\tt rand} function in MATLAB that generates uniformly distributed random numbers between 0 and 1. We then reconstructed the spectrum by calculating the inverse Fourier transform of the ``phase scrambled" Fourier-transformed spectrum. In Figure~\ref{fig:fourier}, we depict an example of calculating the donor fraction when an M0.5 V star was compared to the reconstructed `phase scrambled' spectrum of \kvuma\ in the H-band. 

The advantage of scrambling the phases in Fourier space is that, unlike scrambling the original wavelength-binned data, it works for both white and non-white noise. Any fit done to the ``phase scrambled" spectrum will be spurious since we are fitting signal to pure noise which allows us to place uncertainties on our fits to the original spectrum. We fit the spectra of field stars with  K5 V, K7 V, M0/M0.5 V and M1 V spectral types to each of the phase scrambled spectra and calculated the donor fraction for each wavelength region under investigation listed in Table~\ref{table:fieldstarfits2}. For each field star and each wavelength region under investigation, we repeated this process for 50 randomly generated phase scrambled spectra and examined the variance of the donor fractions. The square root of the variance is then a robust indication of the uncertainty in the donor star fraction. The errors listed with the donor fractions in Table~\ref{table:fieldstarfits2} was obtained from this analysis. Finally, these errors were propagated through the remainder of the steps leading up to the average donor fractions in the J-, H- and K-bands.

\clearpage

\begin{deluxetable}{lccccl}
\tablecaption{Observations of \kvuma\label{table:xteobservations}}
\tablewidth{0pt}
\tablehead{
\colhead {} & \colhead{Date} & \colhead{Individual t$_{exp}$} &  \colhead{Total t$_{obs}$} &\colhead{Orbital Phase Coverage} \\
\colhead{} & \colhead{} & \colhead{(sec)} & \colhead{(hr)} & \colhead{}}
\startdata
Spectroscopy & 04/02/11 & 310 & 2.1 & 0.13 -- 0.39,  0.72 -- 0.98\\
& 04/12/11 & 310 & 2.1 & 0.06 -- 0.32, 0.49 -- 0.76  \\
Photometry & 04/02/11 & 20 & 4.0 & 0.00 -- 0.35, 0.40 -- 0.99 \\
& 04/03/11 &  20 & 3.4\tablenotemark{a} & 0.18 -- 0.99, 0.01 -- 0.04 \\
\enddata
\tablenotetext{a}{Data after the two hours of observation was discarded owing to bad weather.}
\end{deluxetable}

\clearpage

\begin{deluxetable}{ccccccc}
\tabletypesize{\scriptsize}
\tablecaption{Donor star fraction fits\label{table:fieldstarfits2}}
\tablewidth{0pt}
\tablehead{
\colhead{Band} & \colhead{Wavelength range($\mu$m)}&\colhead{Dominant Feature} &  \colhead{$f$ (K5 V)} &   \colhead{$f$ (K7 V)}  &  \colhead{$f$ (M0.5 V)}  &  \colhead{$f$ (M1 V)} }
\startdata
&&&&&& \\
${\bf K}$ &                         ${\bf  2.203-2.214 }   $      &Na I               &0.80$\pm$.15                 &0.70$\pm$.15             &0.65$\pm$.18        &0.62$\pm$.18 \\
&&&&&&\\
\hline
&&&&&&\\
${\bf H}$&    ${\bf 1.48-1.51 }$         &Mg I                   & 0.37$\pm$.07          &  0.37$\pm$.07               & 0.48$\pm$.08         &0.58$\pm$.09\\
&			${\bf 1.568 -1.60}$ 	&Mg I, Si I	      &0.40$\pm$.11           & 0.44$\pm$.09                & 0.47$\pm$.09          &0.63$\pm$.10\\
&	            ${\bf 1.67-1.68}$ 	        &Al I 	             &0.39 $\pm$.12          &  0.39$\pm$.11               & 0.30$\pm$.09          &0.40$\pm$.10\\
&			${\bf  1.70-1.72 }$ 	&Mg I	             & 0.55$\pm$.09         & 0.60$\pm$.09                & 0.65$\pm$.11          &0.70$\pm$.09\\
&&&&&&\\
\hline
&&&&&&\\
${\bf J}$& ${\bf 1.176 - 1.202 }$              & Mg I, Fe I, Si I                      &0.63$\pm$.10                              &0.65$\pm$.13                       & 0.68$\pm$.13          &0.69$\pm$.15\\
&${\bf 1.310-1.317 }$                                & Al I                                        &0.65$\pm$.15                             &0.64$\pm$.16                      & 0.66$\pm$.17           &0.72$\pm$.18\\
&&&&&&\\
\enddata
\end{deluxetable}

\begin{deluxetable}{lcr}
\tablecaption{Adopted Physical Parameters for \kvuma\label{table:parameters}}
\tablewidth{0pt}
\tablehead{
\colhead{Parameter} & \colhead{Value} & \colhead{Reference}}
\startdata
P$_{orb}$ & 0.16995$\pm$0.00012 d & Gonz\'{a}lez Hern\'{a}ndez et al. (2008) \\
K$_{2}$ & 708.8$\pm$1.4 km~s$^{-1}$ & Ibid. \\
q &  0.024$\pm$0.009 & Calvelo et al.\ (2009) \\
$i$ & 68$^{\circ}$ -- 79$^{\circ}$ & This work. \\
M$_{BH}$ & 6.9 -- 8.2 M$_{\odot}$ & This work. \\
Donor spectral type & K7 V -- M1 V & This work. \\
\enddata
\end{deluxetable}

\clearpage

\clearpage

\clearpage

\clearpage

\clearpage

\clearpage

\clearpage

\clearpage

\clearpage

\clearpage

\end{document}